\documentclass[conference]{IEEEtran}
\IEEEoverridecommandlockouts

\usepackage{cite}
\usepackage{amsmath,amssymb,amsfonts}
\usepackage{graphicx}
\usepackage{textcomp}
\usepackage{xcolor}
\def\BibTeX{{\rm B\kern-.05em{\sc i\kern-.025em b}\kern-.08em
    T\kern-.1667em\lower.7ex\hbox{E}\kern-.125emX}}

\makeatletter
\newcommand{\linebreakand}{%
  \end{@IEEEauthorhalign}
  \hfill\mbox{}\par
  \mbox{}\hfill\begin{@IEEEauthorhalign}
  \and
}
\makeatother

\usepackage{subcaption}
\usepackage{algorithm}
\usepackage[noend]{algpseudocode}
\usepackage{multirow}
\usepackage{titlesec}
\usepackage{booktabs}
\usepackage{url}
\usepackage{hyperref}

\newcommand{\algorithmfontsize}{}

\begin{document}

\title{Squire: A General-Purpose Accelerator to Exploit Fine-Grain Parallelism on Dependency-Bound Kernels}

\author{\IEEEauthorblockN{Rubén Langarita}
\IEEEauthorblockA{\textit{Barcelona Supercomputing Center}\\
Barcelona, Spain \\
ruben.langarita@bsc.es}
\and
\IEEEauthorblockN{Jesús Alastruey-Benedé}
\IEEEauthorblockA{\textit{Universidad de Zaragoza}\\
Zaragoza, Spain \\
jalastru@unizar.es}
\and
\IEEEauthorblockN{Pablo Ibáñez-Marín}
\IEEEauthorblockA{\textit{Universidad de Zaragoza}\\
Zaragoza, Spain \\
imarin@unizar.es}

\linebreakand

\IEEEauthorblockN{Santiago Marco-Sola}
\IEEEauthorblockA{\textit{Universitat Politècnica de Catalunya}\\
\textit{Barcelona Supercomputing Center}\\
Barcelona, Spain \\
santiago.marco@bsc.es}
\and
\IEEEauthorblockN{Miquel Moretó}
\IEEEauthorblockA{\textit{Universitat Politècnica de Catalunya}\\
\textit{Barcelona Supercomputing Center}\\
Barcelona, Spain \\
miquel.moreto@bsc.es}
\and
\IEEEauthorblockN{Adrià Armejach}
\IEEEauthorblockA{\textit{Universitat Politècnica de Catalunya}\\
\textit{Barcelona Supercomputing Center}\\
Barcelona, Spain \\
adria.armejach@bsc.es}
}

\maketitle

\begin{abstract}
Multiple HPC applications are often bottlenecked by compute-intensive kernels implementing complex dependency patterns (data-dependency bound).
Traditional general-purpose accelerators struggle to effectively exploit fine-grain parallelism due to limitations in implementing convoluted data-dependency patterns (like SIMD) and overheads due to synchronization and data transfers (like GPGPUs).
In contrast, custom FPGA and ASIC designs offer improved performance and energy efficiency at a high cost in hardware design and programming complexity and often lack the flexibility to process different workloads. 

We propose Squire, a general-purpose accelerator designed to exploit fine-grain parallelism effectively on dependency-bound kernels.
Each Squire accelerator has a set of general-purpose low-power in-order cores that can rapidly communicate among themselves and directly access data from the L2 cache. Our proposal integrates one Squire accelerator per core in a typical multicore system, allowing the acceleration of dependency-bound kernels within parallel tasks with minimal software changes.

As a case study, we evaluate Squire's effectiveness by accelerating five kernels that implement complex dependency patterns.
We use three of these kernels to build an end-to-end read-mapping tool that will be used to evaluate Squire.
Squire obtains speedups up to 7.64$\times$ in dynamic programming kernels. Overall, Squire provides an acceleration for an end-to-end application of 3.66$\times$. 
In addition, Squire reduces energy consumption by up to 56\% with a minimal area overhead of 10.5\% compared to a Neoverse-N1 baseline.
\end{abstract}


\section{Introduction}\label{sec:intro}

Modern multi-core architectures and accelerators have become the cornerstone for accelerating many workloads in scientific computing and engineering~\cite{survey-data-center}. Many efforts have been made to accelerate HPC applications on modern hardware architectures such as CPUs and GPUs, as well as FPGA and custom accelerators (ASICs) for specific workloads~\cite{hpc-survey}. Hence, HPC platforms are increasingly sought after to handle large-scale workloads that exploit different levels of parallelism available in the accelerators.

However, there is an emergent class of workloads that cannot fully exploit the massively parallel capabilities of mainstream accelerators. Many HPC applications are often bottlenecked by the execution of sequential workflows composed of rather small compute-intensive kernels that implement complex dependency patterns (dependency-bound). This is particularly noticeable in life science and healthcare applications, which implement long workflows of data-processing kernels~\cite{survey-genomics-pipeline1}. Often based on stencil and Dynamic Programming (DP) computations, \emph{dependency-bound} kernels tend to be moderate in size and implement complex data-dependency patterns that ultimately restrict parallelism exploitation.

Traditionally, coarse-grain parallelism approaches seek to execute independent kernels concurrently (inter-task) and improve resource utilization~\cite{gnumake,lapack}. However, offloading dependency-bound kernels to specialized hardware accelerators seldom pays off. Due to its rather small size, offloading workloads to decoupled accelerators often incur significant overheads both from data-transfer initialization and movement~\cite{gpu-offload}. Similarly, task-level parallelization approaches often introduce synchronization delays that hinder performance. Also, load imbalance and data sparsity present significant challenges to the flexibility of conventional accelerators, ultimately limiting their performance~\cite{gpu-imbalance}.

Similarly, fine-grain (intra-task) parallelism is difficult to exploit in dependency-bound workloads. Complex dependency patterns and irregular computations make the exploitation of the underlying parallelism challenging~\cite{dp-gpu,gpu-dp2}. Moreover, the need for fine-grain synchronization introduces significant overheads during execution. Ultimately, these kernels often cannot saturate computing resources and effectively exploit fine-grain parallelism~\cite{gpu-dp3,gpu-dp1}.

Mainstream SIMD-based and GPU approaches are limited in accelerating these types of workloads. SIMD approaches face significant limitations in handling sparse data-processing tasks, such as gather and scatter operations, which introduce substantial latency overheads~\cite{simd-gather}. Additionally, SIMD instructions often struggle to implement complex data dependency patterns, limiting their effectiveness in many scenarios. Similarly, GPUs tend to provide only modest performance improvements on these small-scale workloads, as they cannot saturate GPU’s compute resources (i.e., few threads and thread blocks). In addition, GPUs load imbalance and data-transfers overheads limit the performance benefits of offloading these types of kernels~\cite{gpu-offload}.

In contrast, custom FPGA-based and domain-specific ASIC designs offer improved performance and energy efficiency at the cost of complex and expensive hardware design, development, and fabrication processes~\cite{asic-cost}. In some cases, they also suffer from data transfer and synchronization overheads. Ultimately, custom hardware approaches often lack the flexibility to adapt to different workloads and incur significant programming and maintenance costs~\cite{darwin,depgraph,tpu}.

To address these challenges, our \textbf{goal} in this work is to enable (i) efficient exploitation of fine-grain parallelism in dependency-bound kernels, (ii) reducing data-transfers overheads, and (iii) providing hardware support for fast synchronization primitives.

In this work, we propose Squire, a general-purpose accelerator designed to effectively exploit fine-grain parallelism on dependency-bound kernels. Squire is equipped with several general-purpose in-order cores, called workers, and a hardware semaphore for rapid synchronization among the workers. Our proposal incorporates one Squire per core in a typical multicore system, connecting it to the memory hierarchy to directly access the virtual memory space. Each core controls one Squire, rapidly offloading workloads whenever it needs. Since workers share the same Instruction Set Architecture (ISA) as the core, we can develop and compile code for Squire just as we do for the core.

We discuss three potential use cases for our accelerator: data sorting, genomics, and signal processing. We examine some representative kernels and show how fine-grain parallelism can be exploited. These kernels include Radix Sort~\cite{hash-knuth}, Seeding~\cite{minimap2}, Chain~\cite{minimap2}, Smith-Waterman~\cite{sw,sw2}, and Dynamic Time Warping~\cite{dtw}.

This work makes the following contributions.

\begin{itemize}
    \item We propose Squire, a general-purpose accelerator for exploiting fine-grain parallelism on dependency-bound kernels. 
    \item We select five dependency-bound kernels and analyze them. Then, we adapt these kernels to be executed with Squire.
    \item We use three of these kernels to build an end-to-end read-mapping tool that will be used to evaluate Squire.
    \item We show how Squire can speed-up the five evaluated kernels. We also evaluate the end-to-end read-mapper to understand how Squire improves a full application. Finally, we perform an area and energy consumption study.
\end{itemize}

\textbf{Key results.} Squire obtains speedups up to 7.64$\times$ in dynamic programming kernels. Overall, Squire provides an acceleration for an end-to-end application of 3.66$\times$. 
In addition, Squire reduces power consumption by up to 56\% with a minimal area overhead of 10.5\% compared to a Neoverse-N1 baseline.

\section{Motivation} \label{sec:background}

Coarse-grain parallelism is desirable in high-performance computing environments to hide the overheads associated with task management and data movement. This inter-task parallelism means that each processing core operates on an independent task~\cite{minimap2,bwamem2,bwashort,bwalong,bowtie,gem,gnumake,lapack}. However, this approach often struggles to exploit the fine-grain parallelism inherent in many kernels with complex dependencies.

Intra-task fine-grain parallelism is usually tackled via \emph{SIMD} on general-purpose processors or \emph{SIMT} on GPUs. However, these techniques are inefficient when targeting specific algorithms containing dependencies or sparse patterns. Well-known algorithms that suffer such problems include:
Quicksort~\cite{quicksort}, Dynamic Time Warping~\cite{dtw}, and Smith-Waterman~\cite{sw,sw2}. Additionally, data structures with sparse memory patterns, such as FM-Index~\cite{fmindex,fmindexzgz,cofi}, hash tables~\cite{hash-knuth}, and sparse matrix-vector multiplication (SpMV)~\cite{spmv-gpu}, are often limited by the amount of memory-level parallelism that can be exposed. All these patterns are present in many applications.

\begin{figure}[t]
  \centering
  \includegraphics[width=\columnwidth]{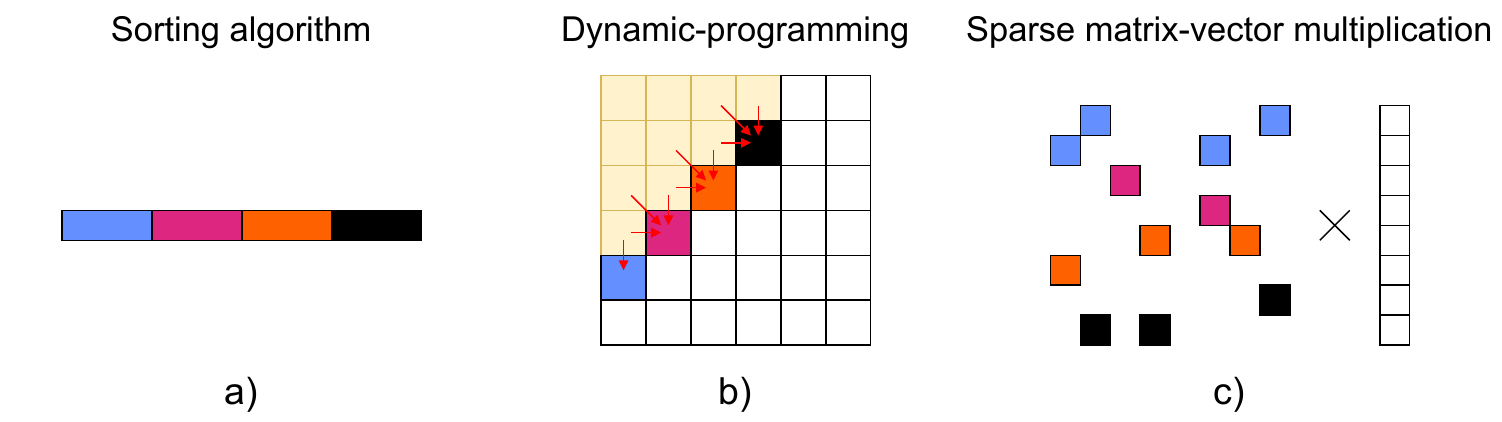}
  \caption{Examples of coarse-grain tasks with fine-grain parallelism: (a) sorting, (b) dynamic programming matrix, (c) sparse matrix-vector multiplication. Each color in the data structures represents a chunk of fine-grain work.}
  \label{fig:nested-parallelism}
\end{figure}

Figure~\ref{fig:nested-parallelism} highlights three kernels that exhibit fine-grain parallelism (shown using different colors) which is challenging to exploit due to existing dependencies. Figures~\ref{fig:nested-parallelism}a and~\ref{fig:nested-parallelism}c show how sorting and SpMV coarse-grain tasks could be further parallelized by processing chunks of the array or independent rows of the matrix in parallel. However, this is not efficient due data-dependent irregular patterns and the fact that SIMD gather/scatter memory operations are not efficient~\cite{simd-gather}.
Subfigure~\ref{fig:nested-parallelism}b shows how parallelism is present per cell in a dynamic programming matrix. For some dynamic programming problems, antidiagonal vectorization is the best way to avoid dependencies; however, it requires data rearranging that diminishes potential gains. Support for exploiting this fine-grain parallelism can benefit a large set of workloads.

GPGPUs have also been proposed to tackle dynamic programming kernels~\cite{dpx,dp-gpu}.
However, GPGPUs are designed for massive parallel workloads, and dependencies and sparsity hinder performance. In addition, offloading fine-grain parallel workloads is not recommended due to the high transfer time, and fine-grain synchronization is also challenging.
For these reasons, GPGPUs obtain modest speed-ups when targeting dynamic programming algorithms~\cite{chain_fpga,chain_gpu,gpu-dp1,gpu-dp2,gpu-dp3}.
Finally, custom hardware solves dependency constraints at the cost of fixing the functionality of the proposed components, losing generality~\cite{genasm,genax,darwin,tpu,depgraph}.

Chain is a dynamic programming algorithm widely used in the field of genomics~\cite{minimap2}.
Lorién et al. show that the SIMD version of chain obtains slowdowns of up to 0.71$\times$ with respect to the scalar version with heuristics~\cite{genarchbench}.
Chain presents dependency-bound patterns in the inner loop, resulting in underutilized vector lanes.
Similarly, Guo et al. show that in the GPU version of chain, 16.3\% of the kernel instructions are control instructions for synchronizing warps~\cite{chain_fpga}.
They use an NVIDIA Tesla P100 for the evaluation~\cite{p100}.
The P100 GPU achieves a 3.17$\times$ speed-up with respect to a 14-core CPU while consuming 300W and occupying 610~mm$^{2}$, resulting in under-utilization of resources.

These limitations highlight the need for a flexible and efficient solution to exploit fine-grain parallelism in dependency-bound workloads. We propose a general-purpose accelerator - Squire - to unlock the parallelism potential of a wide range of workloads while maintaining flexibility and low overhead.

\section{Use cases}\label{sec:analysis}

In this section, we discuss three potential use cases for our accelerator: data sorting, genomics, and signal processing. We examine some representative kernels and demonstrate how fine-grain parallelism can be exploited.

\subsection{Data Sorting}
\label{sec:data_sorting}

Sorting~\cite{hash-knuth} is a widely studied problem in computer science, fundamental to various applications such as search engines, data mining, databases, and numerical methods. The importance of sorting spans from energy-efficient devices~\cite{sort-mobile,sort-mobile2} to GPGPUs and data centers~\cite{sort-gpu,sort-datacenter,sort-energy}.


\textbf{Radix sort}~\cite{hash-knuth} is an efficient sorting algorithm with a time complexity of $O(nk)$, where $n$ is the number of elements in the array and $k$ is the length of the key. This makes it particularly effective to sort arrays of 32-bit or 64-bit integers.
In the first iteration, the algorithm uses the eight most significant bits to divide the array into 2\textsuperscript{8} buckets. This process is repeated recursively for subsequent bits until all bits are processed.

Fine-grain parallelism in Radix Sort can be achieved by dividing the array into smaller chunks, sorting them independently, and merging the results. Previous work has shown how GPGPUs can achieve this using parallel architectures~\cite{radix-merge-finegrain}. We will show that Squire can also leverage this parallelism while experiencing minimal synchronization overhead.


\subsection{Genomics}
\label{sec:chain_background}
\label{sec:seed_background}
\label{sec:sw_background}


In life science research and health care, sequencing technologies have revolutionized the way scientists analyze the genome to uncover biological insights. Modern sequencing technologies can accurately read massive amounts of genome sequences. Afterwards, tools like read mappers are extensively used in multiple sequence data analysis methods to locate (align) the read sequences in a reference genome (e.g., the human genome).
Since alignment algorithms are computationally expensive, read mappers usually follow a seed-and-extend approach. During the seeding stage, the tool searches for partial matches between the query sequence and the reference. These partial matches are hints for the extend stage, where dynamic programming algorithms find the best location for each sequence.


The goal of \textbf{seeding} is to find partial exact matches between an input sequence and the reference genome.
Typically, FM-Index~\cite{fmindex} or hash-tables~\cite{hash-knuth} are used to locate these matches faster.
These data structures are built once and used along all the sequences being aligned, typically several GBs or even TBs of data.

Minimap2~\cite{minimap2} is a well-known read-mapper tool used for long reads (around 10K base pairs).
Initially, Minimap2 builds a hash table that contains the positions in the reference for all combinations of $k$ base pairs (k-mer).
When Minimap2 indexes the hash table with a k-mer, the hash table returns a list of positions in the reference.
To split the sequence into k-mers, Minimap2 establishes a sliding window that moves along the sequence and extracts the lowest k-mer alphabetically in each window.
Each one of these k-mers is called a minimizer.

Minimap2 indexes the hash table with all the minimizers and extracts a list of tuples (called anchors), which consist of the position in the sequence and the position in the reference.
Finally, the anchors are sorted by the position in the reference so that the following stages can easily traverse the list.
For this purpose, Minimap2 uses a radix sort algorithm, which is the most time-consuming step of the entire seeding stage.


\textbf{The Chain kernel}, a 1D dynamic programming algorithm, combines multiple seeds (termed anchors) to create extended matching regions, also called a \emph{chains}.
The algorithm receives a set of sorted anchors and scores each anchor pair based on their proximity and overlap using the following formula:

\begin{equation}
\label{eq:chain}
f(i) = \max\limits_{(i-T) \le j < i} \{ f(j) + \alpha(i,j) - \beta(i,j) \}
\end{equation}

where $f(i)$ is the score of anchor $i$, $\alpha(i,j)$ is a bonus score between the anchors $i$ and $j$, and $\beta(i,j)$ a penalty for gaps and overlaps. Finally, $T$ is the chain iteration threshold.
Notice that the $f(i)$ calculation depends on $f(i-1)$.
On the other hand, the calculation of $\alpha$ and $\beta$ is independent for any $i$ and $j$.

\begin{figure}[t]
  \centering
  \includegraphics[width=0.6\columnwidth]{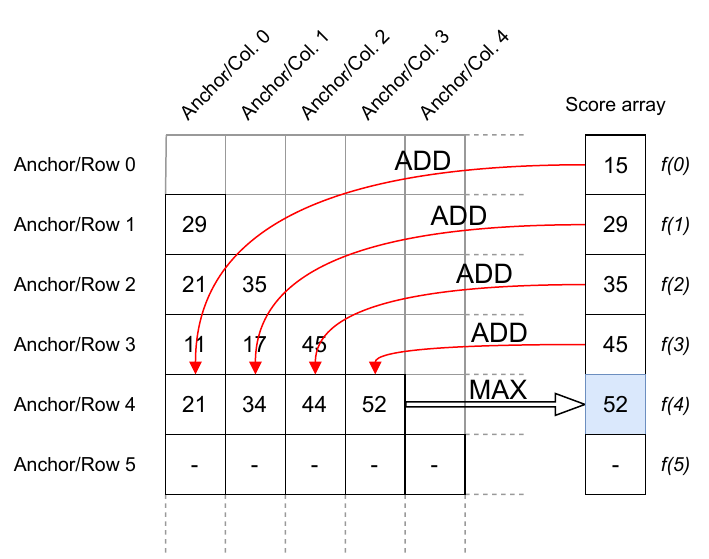}
  \caption{Chain algorithm dependencies. To calculate the score of anchor 4 ($f(4)$), we add all the previous scores to row 4 and perform a maximum. In the next iteration, $f(4)$ will be added to row 5.}
  \label{fig:chain-dependencies}
\end{figure}

Figure~\ref{fig:chain-dependencies} shows the dependencies present in the chain kernel. The score array ($f$ in Equation~\ref{eq:chain}) is added to the match-up scores ($\alpha$ and $\beta$ in Equation~\ref{eq:chain}).
Then, the maximum of the row is calculated, and the score of anchor 4 is obtained.
Notice that $f(4)$ will be added to subsequent rows, thus creating a dependency in the outer loop.

In addition to the score array, a predecessor array is computed, which stores the index of the best match-up for each anchor. Once both arrays are computed, a backtracking process is performed. Starting from the best score in the score array, the algorithm recursively traces back through its predecessors. The resulting list of predecessors forms a \textit{chain}.

Dynamic programming 1D algorithms typically involve an inner loop with dependencies across iterations, which limits the amount of parallelism that can be exploited. The fine-grain parallelism arises from the computation of elements that depend on previously computed iterations. As illustrated in Figure~\ref{fig:chain-dependencies}, the element $f(4)$ in the score array requires all elements smaller than $4$. However, it is unnecessary to wait for $f(3)$ to complete before calculating certain cells in the row. For example, we can compute cell $x$ in row 4 as soon as $f(x)$ is available by dynamically checking the computed scores. A system of detached computing elements could facilitate such computations effectively.

\textbf{Smith-Waterman} is a 2D dynamic programming algorithm used during the extend stage for sequence alignment~\cite{sw,sw2}.
As this paper discusses another 2D dynamic programming algorithm, Dynamic Time Warping, in detail later, we do not elaborate further on Smith-Waterman here.

\subsection{Signal Processing} \label{sec:signal_processing_background}

Signal processing focuses on analyzing various types of signals, including sound, images, potential fields, seismic data, altimetry, and scientific measurements. A signal represents a flow of information originating from a source, which can take many forms, such as mechanical, optical, magnetic, electrical, or acoustic. Signals can be digital, characterized by discrete values, such as semaphores, Morse code, or the contents of computer memory. Conversely, signals can also be analog, encompassing continuous values like pressure, temperature, or velocity.


\textbf{Dynamic Time Warping} (DTW) is a 2D dynamic programming algorithm designed to align two signals to measure their similarity. It is widely used in applications such as speech recognition, speaker recognition, and music recognition. DTW constructs a dynamic programming matrix with a computational complexity of $O(n \times m)$, where $n$ and $m$ represent the lengths of the signals being aligned.

\begin{figure}[t]
  \centering
  \includegraphics[width=0.5\columnwidth]{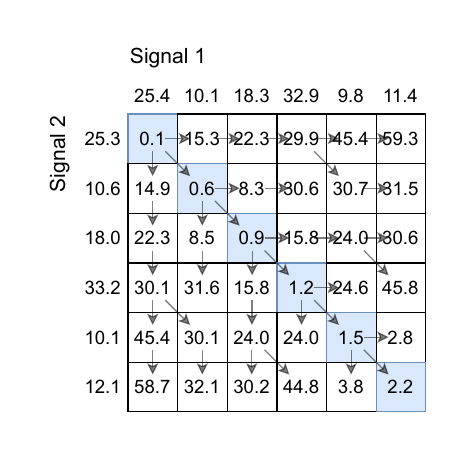}
  \caption{DTW matrix representation. The samples from signal 1 are set at the top of the matrix, while samples from signal 2 are set at the left. For each cell DTW computes Equation~\ref{eq:dtw}. Arrows indicates the minimum value from Equation~\ref{eq:dtw}.}
  \label{fig:dtw-matrix}
\end{figure}

Figure~\ref{fig:dtw-matrix} provides an example of a DTW matrix. Each cell in the matrix is computed using the following equation:

\begin{equation}
\label{eq:dtw}
\begin{split}
M[i,j] = abs(S[i]-R[j]) + \min \{ &M[i-1,j-1] , \\
&M[i-1,j] , \\
&M[i,j-1] \}
\end{split}
\end{equation}

Where $M[i,j]$ is the cell in row $i$ and column $j$, while $S$ and $R$ are the aligned signals.
The equation calculates the value of the cell $[i,j]$ as the minimum value of the left, top, and left-top cells plus the absolute difference between $S[i]$ and $R[j]$.
This dependency on the left, top, and left-top cells complicates parallelization.

Similarly, the aforementioned Smith-Waterman algorithm also exhibits the same dependency patterns as DTW. Like DTW, Smith-Waterman also constructs a matrix with dependencies on the left, top, and left-top cells.


Typically, fine-grain parallelism for 2D dynamic programming kernels is primarily achieved using SIMD techniques. The most common approaches are anti-diagonal vectorization~\cite{wozniak-antidiagonal} and inter-task vectorization~\cite{rognes2-intertask}. Both methods require prior reorganization of the data structures and often suffer from under-utilization of SIMD lanes.

On one hand, SIMD enforces a lockstep execution order, limiting flexibility. On the other hand, dividing work in this manner can lead to load imbalance. A potential solution to these issues is to increase the size of the work chunks assigned to each computing element. As shown in Section~\ref{sec:squirization_dtw}, the dynamic programming (DP) matrix could be divided by columns and distributed among the computing elements. To implement this effectively, we would require a set of independent computing elements capable of asynchronously processing their assigned chunks of work.

\subsection{Discussion}

\begin{figure*}[t]
    \centering
    \includegraphics[width=\textwidth]{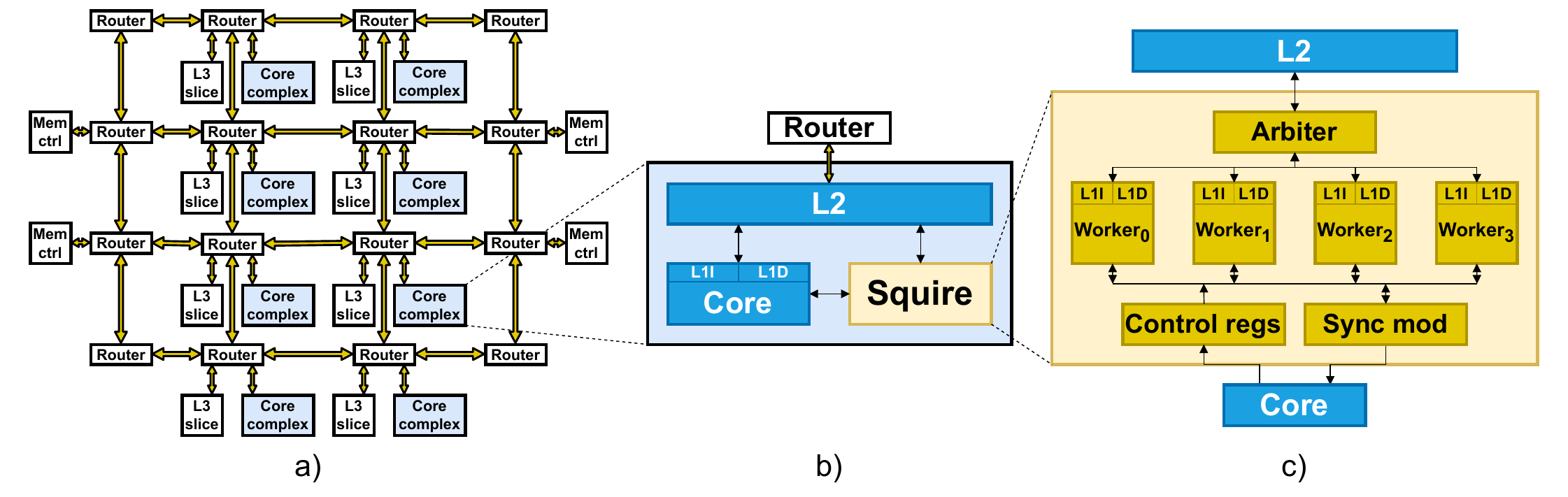}
    \caption{Squire architectural overview: (a) the simulated multi-core system with a 4x4 NoC and four memory controllers, where each central router holds a core complex and one slice of the L3 cache; (b) a core complex contains one OoO core with private L1 caches, a private L2 cache, and a Squire; (c) Squire contains a set of workers, several control registers, a synchronization module, and an arbiter to communicate with the memory hierarchy (L2).}
    \label{fig:engine_scheme}
\end{figure*}

We have demonstrated that fine-grain parallelism exists in several key dependency-bound kernels. Exploiting this parallelism effectively requires a set of generic, independent, general-purpose computing elements. Such a system must include a mechanism for rapid work offloading and efficient synchronization among the computing elements.
Building on these principles, the next section introduces Squire, a general-purpose accelerator designed specifically to address the challenges of dependency-bound fine-grain parallelism.

\section{Squire}\label{sec:accelerator}

One of the main goals of Squire is to make it general-purpose, enabling workloads with inherent fine-grain parallelism to benefit from the accelerator. To achieve this, we identify two key design principles: (i) enable low-latency execution of fine-grain tasks and (ii) provide architectural support for fast synchronization between processing units to manage dependencies. Hence, our hardware accelerator must have the following features:

\begin{itemize}
    \item A set of general-purpose processing units sharing a unified memory view with the host core.
    \item A synchronization mechanism to enable rapid communication among processing units.
\end{itemize}

\subsection{Squire Design}\label{sec:squire_overview} \label{sec:worker_design}

Figure~\ref{fig:engine_scheme} shows the architectural overview of Squire, a general-purpose accelerator for dependency-bound fine-grain parallelism. Figure~\ref{fig:engine_scheme}a illustrates a conventional multi-core SoC with a distributed L3 cache, where each core complex contains two levels of private caches. Figure~\ref{fig:engine_scheme}b depicts the integration of Squire into the system, where each core complex is augmented with a Squire block interfaced with the private L2 cache. Finally, Figure~\ref{fig:engine_scheme}c shows that Squire consists of a set of very simple general-purpose in-order cores, termed \emph{workers}. In addition, Squire features control registers and a synchronization module, which are visible to both the host core and the workers.

Typically, hardware accelerators solve dependency-bound parallelism, such as dynamic programming, using systolic arrays~\cite{chain_fpga,darwin,genasm}. However, these solutions rely on hard-wired components that serve a fixed purpose and usually target a specific kernel.
In order to increase the flexibility of Squire, we propose employing simple in-order cores with small area and power consumption requirements for each worker. To simplify the design, we assume these cores share the same base ISA as the host core. In addition, each worker has small, private data and instruction caches. We define the size of these caches with a design space study in Section~\ref{sec:cache-exploration}.

A host core can offload computation to the workers via a simple API (see Section~\ref{sec:squire-api}) that sets a function's address and the necessary arguments into the control registers. Then, the workers start executing the workload using regular instructions. If the host core has recently accessed the input data, it is likely to still reside in the L2 cache, reducing data transfer latency.

To orchestrate L2 access requests from the worker cores, we employ a shared bus coupled with a centralized arbiter. The arbiter selects one request per cycle from the set of pending L2 accesses issued by the workers. This design enforces a single L2 access per cycle, thereby requiring only a single extra read/write port on the L2 cache, reducing implementation complexity. Cache coherence is maintained through a snoop-based protocol, where all workers monitor the L2 bus for invalidation messages. This is practical given the simplicity of the in-order cores. Moreover, workers are designed to target workloads that maximize L1 data reuse. Empirically, even with 32 workers active, the system sustains an average of no more than one L2 access every two cycles, demonstrating that accessing the L2 cache is not a primary bottleneck.






Tasks are distributed using traditional coarse-grain parallelism, with OpenMP assigning independent workloads to host cores~\cite{openmp}. In read mapping tools, for example, each host core aligns a subset of sequences. These tasks are typically dependency-bound, limiting the effectiveness of SIMD and instruction-level parallelism. Squire addresses this by subdividing tasks into fine-grain sub-tasks, enabling nested parallelism even in dependency-bound kernels.

\subsection{Synchronizing Workers} \label{sec:sync-workers}

The synchronization mechanism is used to coordinate the workers, and it is visible to the host core as well as the workers. We have designed the mechanism to enable modeling dependencies for two distinct common use cases.

\begin{table}[t]
\centering
\caption{Squire programming interface. For each API call, we provide a brief description and who can use the API call.}
\label{tab:api_squire}
\resizebox{\columnwidth}{!}{%
\begin{tabular}{@{}lll@{}}
\toprule
\textbf{API call} & \textbf{Description} & \textbf{Caller}  \\ \midrule
\multirow{2}{*}{\texttt{start\_squire(f,a)}} & Squire executes \texttt{f} function with \texttt{a} arguments. & \multirow{2}{*}{Core} \\
 & \textit{Counters} reset to 0. & \\
\texttt{stop\_worker()} & Suspends the worker execution. & Workers \\
\texttt{id\_worker()} & Returns the worker ID. & Workers \\
\texttt{num\_workers()} & Returns the total number of workers. & Core/Workers \\
\texttt{inc\_lcounter(w)} & Increments the \textit{local counter} \texttt{w} by one. & Workers \\
\texttt{inc\_gcounter()} & Increments the \textit{global counter} by one. & Workers \\
\texttt{wait\_lcounter(w,s)} & Waits until the \textit{local counter} \texttt{w} is greater or equal to \texttt{s}. & Core/Workers \\\texttt{wait\_gcounter(s)} & Waits until the \textit{global counter} is greater or equal to \texttt{s}. & Core/Workers \\ \bottomrule
\end{tabular}%
}
\end{table}

On the one hand, our aim is to tackle algorithms that perform computation over 1D data structures. To achieve this, we use a simple mechanism that features a hardware atomic counter, referred to as \textit{global counter}. This will enable handling loops where iteration \textit{i} conditionally consumes the data produced by iteration \textit{i-1}.
For this purpose, we require the workers to increment the \textit{global counter} in order, i.e., if worker \textit{x} increments the \textit{global counter} before worker \textit{x-1}, the increment is saved in a structure until worker \textit{x-1} increments the \textit{global counter}.
To implement this with a non-blocking scheme, we instantiate one queue per worker and a token.
The token indicates which worker is the next to increment the \textit{global counter} and is initialized to zero.
If worker \textit{x} wants to increment the \textit{global counter} and the token contains the value \textit{x-1}, an increment request is enqueued in \textit{x}'s queue.
When worker \textit{x-1} increments the \textit{global counter}, the queues are searched for pending increments in order, and the token is updated accordingly.

On the other hand, we want to efficiently handle workloads with 2D data structures, such as dynamic programming matrices with vertical and horizontal dependencies.
For this reason, we also instantiate an array of hardware atomic counters, with a length equal to the number of workers, referred to as \textit{local counters}.
When filling a dynamic programming matrix, a worker increments its \textit{local counter} each time it computes a matrix row.
Thus, worker \textit{x} can check \textit{local counter} \textit{x-1} before starting the next row.

This set of hardware atomic counters is implemented as 64-bit registers that can be accessed in one cycle.

\subsection{Squire API} \label{sec:squire-api}

Table~\ref{tab:api_squire} describes the Squire programming interface. Each functionality in the table defines a new ISA primitive to interact with the accelerator. The table also specifies whether the host core, the workers, or both can invoke the primitives.
Section~\ref{sec:radix_opt} shows how the Squire API works using radix sort as an example.

\section{Using Squire}\label{sec:software}

This section shows how to use Squire for several kernels.
We describe the implementation process for the Radix Sort, Chain, and DTW algorithms in Squire.
Finally, we discuss some alternatives considered for certain implementation details.

\subsection{Sorting: Radix Sort} \label{sec:radix_opt}

\begin{algorithm}[t]
\caption{Radix Sort Squire version}
\label{alg:radix_squire}
\begin{algorithmic}[1]
\algorithmfontsize
\Function{RADIX}{X[N]} \label{alg:radix_ooo}
\If {N $>$ 10000} \label{alg:radix_check_anchors}
\State{start\_squire(\textsc{RADIX\_Workers}, X) \label{alg:radix_start_squire}}
\State{wait\_gcounter(num\_workers())} \label{alg:radix_ooo_wait}
\State{MERGE\_SORTED\_ARRAYS(X) \label{alg:radix_merge}}
\Else
\State{RADIX\_KERNEL(X[0:N]) \label{alg:radix_ooo_kernel}}
\EndIf
\EndFunction
\Function{RADIX\_Workers}{X[N]} \label{alg:radix_workers}
\State{start = id\_worker() $\times$ (N / num\_workers()) \label{alg:radix_chunk1}}
\State{end = (id\_worker()+1) $\times$ (N / num\_workers()) \label{alg:radix_chunk2}}
\State{RADIX\_KERNEL(X[start:end]) \label{alg:radix_workers_kernel}}
\State{inc\_gcounter() \label{alg:radix_inc_semaphore}}
\State{stop\_worker() \label{alg:radix_stop_worker}}
\EndFunction
\end{algorithmic}
\end{algorithm}

The pseudocode for Squire’s radix sort implementation is shown in Algorithm~\ref{alg:radix_squire}. The host core executes the \texttt{RADIX} function (Line~\ref{alg:radix_ooo}), which calls \texttt{start\_squire} with the function and input data addresses as arguments (Line~\ref{alg:radix_start_squire}). This call writes the addresses to Squire’s control registers, sets the workers’ program counters to the function’s entry point, and resets internal counters. The workers then execute the \texttt{RADIX\_Workers} function (Line~\ref{alg:radix_workers}). Each worker retrieves its ID and the total number of workers via the \texttt{id\_worker} and \texttt{num\_workers} APIs, using this information to evenly partition the input array (Lines~\ref{alg:radix_chunk1}–\ref{alg:radix_chunk2}). Each chunk is sorted using the standard radix sort algorithm (Line~\ref{alg:radix_workers_kernel}). Upon completing its chunk, a worker increments the global counter and halts (Lines~\ref{alg:radix_inc_semaphore}–\ref{alg:radix_stop_worker}). Meanwhile, the host core waits until the global counter matches the number of workers (Line~\ref{alg:radix_ooo_wait}). At this point, the input has been divided into n sorted subarrays, which the host core merges using a min-heap (Line~\ref{alg:radix_merge}).
Squire may not be beneficial when the workload is too small. To address this, Algorithm~\ref{alg:radix_squire} includes a check to ensure that at least 10,000 elements are present before activating Squire (Line~\ref{alg:radix_check_anchors}); otherwise, the host core handles sorting directly (Line~\ref{alg:radix_ooo_kernel}).


\subsection{1D Dynamic Programming: Chain} \label{sec:chain-squire}

We describe the process of integrating the Chain kernel into Squire.
First, we show the pseudocode for the baseline version of the Chain kernel.
Next, we outline generic software modifications to enable parallelism, and finally, we show the necessary changes to integrate Chain into Squire.

\subsubsection{Baseline Chain Kernel}

\begin{algorithm}[t]
\caption{Chain kernel baseline version}
\label{alg:baseline}
\begin{algorithmic}[1]
\algorithmfontsize
\Function{CHAIN}{A[N]} \label{alg:chain_function} \Comment{A: anchors array}
\State{T = 5000 \label{alg:tassign_base}}
\For{i = 0; i $<$ N; i++} \label{alg:outter_o}
    \For{j = i-1; j $\ge$ i-T; j-\hspace{0pt}-} \label{alg:inner_o}
        \State{AUX[j] = $\alpha$(A[i], A[j]) - $\beta$(A[i], A[j]) \label{alg:alpha}}
        \State{AUX[j] += F[j] \label{alg:add_score} \Comment{Consume \texttt{F[j]}}}
    \EndFor
\State{F[i] = MAX(AUX) \label{alg:update_score} \Comment{Generate \texttt{F[i]}}}
\EndFor
\EndFunction
\end{algorithmic}
\end{algorithm}

Algorithm~\ref{alg:baseline} shows the pseudocode for the original chain kernel.
The function \texttt{CHAIN} receives an array of anchors sorted by position in the reference (Line~\ref{alg:chain_function}).
The kernel consists of two nested loops.
The outer loop (Line~\ref{alg:outter_o}) goes through all the anchors sorted by reference position, while the inner loop (Line~\ref{alg:inner_o}) iterates through the \texttt{T} anchors prior to anchor \texttt{i} and performs a \emph{match-up} between each of them and anchor \texttt{i}.
Line~\ref{alg:alpha} corresponds to the calculation of $\alpha$ and $\beta$ in Equation~\ref{eq:chain}, and Line~\ref{alg:add_score} to the addition of $f(j)$ in Equation~\ref{eq:chain}.
Line~\ref{alg:update_score} performs the maximum, obtaining $f(i)$.
Notice that Line~\ref{alg:alpha} can be computed in parallel for all the anchors, while Line~\ref{alg:add_score} must wait for the generation of \texttt{F[j]} by Line~\ref{alg:update_score}.

\subsubsection{Enabling Fine-Grain Parallelism} \label{sec:chain_thres}


To delay the consumption of \texttt{F[i]} from Line~\ref{alg:update_score} to Line~\ref{alg:add_score}, we alter the order of the inner loop (Line~\ref{alg:inner_o}).
To achieve this, we traverse the anchors in reverse order, i.e., from \texttt{i-T} to \texttt{i-1}.
In addition, to isolate dependency-free parallelism from the dependencies imposed by Line~\ref{alg:add_score}, we fission the inner loop (Line~\ref{alg:inner_o}), effectively detaching the computation of $\alpha$ and $\beta$ in Line~\ref{alg:alpha} from the addition in Line~\ref{alg:add_score}.


By default, the chain algorithm has a threshold on the number of anchors it visits backward (\texttt{T} in Lines \ref{alg:tassign_base} and \ref{alg:inner_o}). However, the best match-up is typically found during the initial iterations. In addition, the chain implements some heuristics to stop the inner loop earlier. For example, if the match-up scores are below a threshold.
Therefore, the chain kernel visits fewer anchors, and typically only the first few are useful.
Consequently, we can reduce \texttt{T} with a negligible penalization in accuracy. We observe a misprediction rate lower than 9 per million when setting \texttt{T} to 64. Therefore, we use this value for our final evaluation.
Although the overall Minimap2 accuracy remains almost unchanged, limiting \texttt{T} skips some match-ups, and some computation shifts to the align stage.


In the original implementation, after performing the chain stage, there is a second opportunity to rerun the chain algorithm if the area covered by the anchors is not large enough.
The chain kernel is executed again with looser parameters and simpler versions of $\alpha$ and $\beta$ functions.
We modify the second chain run to use the same function as in the first chain run while applying the new parameters.
This simplifies the implementation process while preserving the output of the original algorithm. As a summary of the software modifications:

\begin{itemize}
    \item We have reversed the order in which we traverse the inner loop (Line~\ref{alg:inner_o}).
    \item We fission the inner loop. Line~\ref{alg:alpha} will be executed in the first loop, and Line~\ref{alg:add_score} in the second one.
    \item We limit the number of anchors visited backward to 64 (\texttt{T} in Lines \ref{alg:tassign_base} and \ref{alg:inner_o}).
    \item We reformulate the second chain run to use the same function as in the first run.
\end{itemize}

Algorithm~\ref{alg:workers} shows the modified code with all these changes.

\subsubsection{Squire Integration} \label{sec:chain_parallelism}

\begin{algorithm}[t]
\caption{Chain kernel Squire version}
\label{alg:workers}
\begin{algorithmic}[1]
\algorithmfontsize
\Function{CHAIN\_Workers}{A[N]} \Comment{A: Anchors array}
\State{T = 64 \label{alg:tassign_squire}}
\For{i = id\_worker(); i $<$ N; i += num\_workers()} \label{alg:outter1_w}
    \For{j = i-T; j $\le$ i-1; j++} \label{alg:inner1_w}
        \State{AUX[j] = $\alpha$(A[i], A[j]) - $\beta$(A[i], A[j]) \label{alg:alpha_w}}
    \EndFor
    \For{j = i-T; j $\le$ i-1; j++} \label{alg:inner2_w}
        \If{AUX[j] $\neq$ -$\infty$} \label{alg:ifbeta}
            \State{wait\_gcounter(j+1) \label{alg:semaphore_check}}
            \State{AUX[j] += F[j] \label{alg:add_score_w} \Comment{Consume \texttt{F[j]}}}
        \EndIf
    \EndFor
    \State{F[i] = MAX(AUX) \label{alg:fi_calculation}} \Comment{Generate \texttt{F[i]}}
    \State{inc\_gcounter() \label{alg:semaphore_add}}
\EndFor
\State{stop\_worker()}
\EndFunction
\end{algorithmic}
\end{algorithm}

Algorithm~\ref{alg:workers} shows the modified chain kernel adapted for Squire. The work is divided in a round-robin fashion (Line~\ref{alg:outter1_w}); e.g., with four workers, worker 0 computes the scores of anchors (0, 4, 8, ...), worker 1 computes the scores of anchors (1, 5, 9, ...), and so on.

Note that now the loop in Line~\ref{alg:inner1_w} can be computed in parallel without dependencies using the workers. Once all the $\alpha$ and $\beta$ values have been computed, the second loop in Line~\ref{alg:inner2_w} proceeds to compute the remaining part, which has dependencies across workers (red lines in Figure~\ref{fig:chain-dependencies}). The dependencies are expressed by waiting on the \textit{global counter} until it contains the desired value (Line~\ref{alg:semaphore_check}), which means the dependent \texttt{F[j]} has been computed. Once the current \texttt{F[i]} is computed (Line~\ref{alg:fi_calculation}), we increment the \textit{global counter} to notify (Line~\ref{alg:semaphore_add}) the consumers of that value.

When $\beta$ (the penalization score) is high enough, we can stop the computation for that match-up.
For these cases, we add a conditional statement (Line~\ref{alg:ifbeta}).
Note that bypassing the \texttt{wait\_gcounter} instruction could cause a race condition.
For this purpose, we have implemented the mechanism described in Section~\ref{sec:sync-workers}, where we enforce the order of the increments in the \textit{global counter}.

\subsection{2D Dynamic Programming: DTW}
\label{sec:squirization_dtw}

\begin{figure}[t]
  \centering
  \includegraphics[width=0.5\columnwidth]{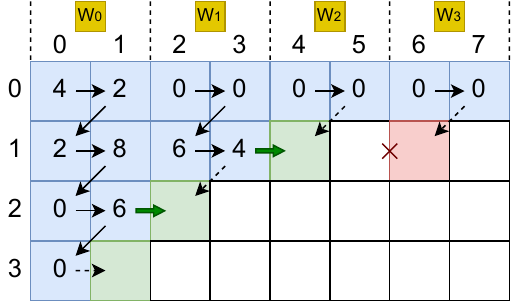}
  \caption{DTW work distribution among workers. Worker 0 (W\textsubscript{0}) computes columns 0 and 1, worker 1 (W\textsubscript{1}) columns 2 and 3, worker 2 (W\textsubscript{2}) columns 4 and 5, and worker 3 (W\textsubscript{3}) columns 6 and 7. The workers compute the cells following the path indicated by the arrows.}
  \label{fig:sw-squire}
\end{figure}

We now detail how to use Squire to exploit fine-grain parallelism in DTW.
Other well-known 2D DP kernels (e.g., Smith-Waterman, Needleman-Wunsch, etc.) exhibit the same patterns when computing the DP matrix.

Figure~\ref{fig:sw-squire} shows a graphical scheme of how Squire would compute the DTW matrix.
A set of consecutive columns is assigned to each worker; worker 0 (W\textsubscript{0}) computes columns 0 and 1, worker 1 (W\textsubscript{1}) columns 2 and 3, and so on.
Each cell \textit{(i,j)} of the matrix has a dependency with cells \textit{(i-1,j)}, \textit{(i,j-1)} and \textit{(i-1,j-1)}.
The workers compute their columns in a row-wise order. Hence, they do not have to worry about the vertical and diagonal dependencies.
To solve horizontal dependencies at the boundaries, the \textit{local counters} from the \textit{synchronization module} are used.

Algorithm~\ref{alg:dtw-workers} shows the pseudocode for the Squire version of DTW.
First, the work is evenly divided among the workers (Lines~\ref{alg:dtw-start} and \ref{alg:dtw-end}).
The outer loop iterates through the rows (Line~\ref{alg:dtw-outter}), while the inner loop iterates through the assigned columns of the corresponding worker (Line~\ref{alg:dtw-inner}).
The equations of DTW are implemented in Lines \ref{alg:dtw-min}, \ref{alg:dtw-cost}, and \ref{alg:dtw-cell}.
To synchronize workers at the boundaries, worker~\textit{x} increments the \textit{local counter} \textit{x} when it finishes a row (Line~\ref{alg:dtw-inccounter}), so worker~\textit{x+1} knows the dependency for that row is solved.
Similarly, when worker~\textit{x} starts a row, it waits for worker~\textit{x-1} to finish its chunk of the row (Line~\ref{alg:dtw-waitcounter}).
Note that worker~0 has no horizontal dependencies. Therefore, it skips the synchronization (Line~\ref{alg:dtw-ifidworker}).

\begin{algorithm}[t]
\caption{DTW kernel Squire version}
\label{alg:dtw-workers}
\begin{algorithmic}[1]
\algorithmfontsize
\Function{DTW\_Workers}{A[N], B[M]}
\State{start = id\_worker() $\times$ (M / num\_workers())} \label{alg:dtw-start}
\State{end = (id\_worker() + 1) $\times$ (M / num\_workers())} \label{alg:dtw-end}
\For{i = 0; i $<$ N; i++} \label{alg:dtw-outter}
    \If{id\_worker() $\neq$ 0} \label{alg:dtw-ifidworker}
        \State{wait\_lcounter(id\_worker()-1, i+1)} \label{alg:dtw-waitcounter}
    \EndIf
    \For{j = start; j $<$ end; j++} \label{alg:dtw-inner}
        \State{PREV $\leftarrow$ MIN(M[i-1,j], M[i,j-1], M[i-1,j-1])} \label{alg:dtw-min}
        \State{COST $\leftarrow$ COST\_FUNC(A[i], B[j])} \label{alg:dtw-cost}
        \State{M[i,j] $\leftarrow$ PREV + COST} \label{alg:dtw-cell}
    \EndFor
    \State{inc\_lcounter(id\_worker())} \label{alg:dtw-inccounter}
\EndFor
\State{stop\_worker()} \label{alg:dtw-stopworker}
\EndFunction
\end{algorithmic}
\end{algorithm}

\subsection{Discussion}

Throughout the development of Squire, we have examined several ideas regarding certain implementation details.

For communication among the workers, we have considered message-passing through a crossbar, a FIFO, or a ring.
Finally, we have used the shared L2 cache since the worker's messages are part of the output, avoiding the need to write the same data twice.

We also considered other synchronization mechanisms besides the \textit{counters}.
Initially, the message-passing mechanism would be used as the synchronization point.
We explored expanding the \textit{synchronization module} functionality, allowing subtractions and arbitrary additions over the counter.
The current \textit{synchronization module} specifications are sufficient for the algorithms we use, but they could be extended in the future.

Finally, as we explained in Section~\ref{sec:squire_overview}, workers must increase the \textit{global counter} in order.
We considered solving this problem in software by waiting for the \textit{global counter} to reach its correct value before incrementing it, e.g., in Algorithm~\ref{alg:workers} adding \texttt{wait\_gcounter(i)} between Lines \ref{alg:fi_calculation} and \ref{alg:semaphore_add}.
However, this approach would harm available parallelism and performance.

\section{Evaluation Methodology}\label{sec:methodology}

\subsection{Architectural Simulation} \label{sec:gem5_setup}

We prototype Squire using the gem5 simulator v23.0~\cite{gem5,gem52}.
We simulate a multicore system consisting of 8 Neoverse-N1-like out-of-order cores, three levels of cache, 4 HBM2e memory channels, and a mesh-based network-on-chip modeling the AMBA 5 CHI protocol, as shown in Figure~\ref{fig:engine_scheme}a. Each host core features a Squire engine that faithfully models the described architecture. The simulated system runs Ubuntu 22.04 with Linux kernel 5.4.65. Table~\ref{tab:ooo_sizes} summarizes the architectural parameters.

\begin{table}[t]
\caption{Simulated architectural parameters.}
\resizebox{\columnwidth}{!}{%
\begin{tabular}{@{}ll@{}}
\toprule
\textbf{Cores} & 8 Neoverse-N1-like Armv8 out-of-order cores 2.4 GHz \\
\textbf{Structure entries} & ROB: 224 $|$ LD/ST queues: 96/96 $|$ Inst. queue: 120 \\ \midrule
\textbf{OoO Private L1 I\&D} & 64 KB, 4-way, 1 cycle data access, 32 MSHRs \\
\textbf{Private L2} & 512 KB, 8-way, 4 cycle data access, 64 MSHRs \\
\textbf{Shared L3} & \multirow{2}{*}{\begin{tabular}[c]{@{}l@{}}Mostly exclusive, 8 slices of 1 MB, 16-way,\\ 10 cycles data access, 128 MSHRs\end{tabular}} \\
 &  \\ \midrule
\textbf{Coherence protocol} & MOESI-like AMBA 5 CHI specification \\
\textbf{Network topology} & 4$\times$4 2D mesh, 1 cycle routers, 1 cycle links (Fig.~\ref{fig:engine_scheme}a) \\ \midrule
\textbf{Memory} & 1 HBM2 stack, 300 GB/s \\ \midrule
\textbf{Worker} & \multirow{2}{*}{\begin{tabular}[c]{@{}l@{}}Cortex-M35P-like Armv8 4-stage dual-issue\\ in-order cores 2.4 GHz \end{tabular}} \\
 &  \\ \bottomrule
\end{tabular}%
}
\label{tab:ooo_sizes}
\end{table}

\begin{table}[t]
\caption{Size of the datasets used in the evaluation.}
\resizebox{\columnwidth}{!}{%
\begin{tabular}{@{}rccccc@{}}
\toprule
 & \textbf{RADIX} & \textbf{SEED} & \textbf{CHAIN} & \textbf{SW} & \textbf{DTW} \\
\midrule
\textbf{\# experiments} & 15 & 5 & 5 & 5 & 2 \\
\textbf{\# inputs/exp.} & 8 arrays & 24 seq. & 24 arrays & 6195 align. & 5000 align. \\
\textbf{Input avg. size} & 53536 elems. & 23014 bps & 53536 anchors & 1373 bps & 221 samples \\
\textbf{Size st. dev.} & 36886 & 15075 & 36886 & 2950 & 101 \\
\textbf{Mem. footprint} & 837 KB & 22.5 KB & 837 KB & 3.27 KB & 1.72 KB \\
\bottomrule
\end{tabular}%
}
\label{tab:kernel_inputs}
\end{table}

\subsection{Workloads and Inputs}\label{sec:methodology_kernels}

Table~\ref{tab:kernel_inputs} details the inputs used for each kernel.
All the inputs have been extracted from real genomics and signal-processing datasets.
To evaluate Squire, we use the five kernels described below.

\textbf{Radix Sort (RADIX)} Radix sort is shown in Section~\ref{sec:data_sorting}.
We did 15 experiments.
In each experiment, we sort eight arrays, one for each out-of-order core.
Some of the arrays used for radix sort have less than 10,000 elements, thus avoiding offloading work to Squire (see Section~\ref{sec:radix_opt}).
We divide the array into equal chunks and use Squire to sort them (see Section~\ref{sec:radix_opt}).

\textbf{Seeding (SEED)} We evaluate the seeding algorithm from Minimap2~\cite{minimap2} (see Section~\ref{sec:seed_background}).
We use five input sequences datasets (see Table~\ref{tab:perf_inputs}).
Each one of the datasets has 24 sequences, hence, each out-of-order core performs three seeding processes.
The most consuming part of seeding is the final sorting of the seeds. Therefore, we use the Squire version of the radix sort algorithm explained above.

\textbf{Chain (CHAIN)} Chain is a dynamic algorithm used in Minimap2~\cite{minimap2} (see Section~\ref{sec:chain_background}).
As in seeding kernel, we use five input sequences datasets, where each one has 24 sequences, resulting in three chain processes per out-of-order. 
The anchors are assigned to the workers in a round robin manner (see Section~\ref{sec:chain-squire}).

\textbf{Smith-Waterman (SW)} Smith-Waterman is a 2D dynamic programming algorithm used for aligning (see Section~\ref{sec:sw_background}).
We use the same datasets used in seeding and chain.
These datasets produce several alignments that we use as inputs in Smith-Waterman.
The work has been distributed using the same approach as for DTW (see Section~\ref{sec:squirization_dtw}).

\textbf{Dynamic Time Warping (DTW)} Dynamic Time Warping is a 2D dynamic programming algorithm (explained in Section~\ref{sec:signal_processing_background}) used for signal processing.
We use two synthetic datasets of 5,000 alignments of floating point numbers.
The small dataset has an average alignment size of 133 samples, while the larger one has 380 samples on average.
Each worker is the responsible of computing several contiguous columns (see Section~\ref{sec:squirization_dtw}).

\subsection{Evaluation of an End-to-End Read-Mapping Application} \label{sec:inputs}\label{sec:end2end_eval}

With the kernels introduced above, we have built an end-to-end read-mapping tool that receives a set of sequences and produces alignments.
We use Minimap2~\cite{minimap2} as the skeleton for our read-mapper since two of the evaluated kernels are extracted from Minimap2 (SEED and CHAIN).
We combine SEED, CHAIN, and SW into a single application to set up a read-mapper that serves as a test-bench for evaluating the speed-up achieved on an end-to-end application when using Squire.

Table~\ref{tab:perf_inputs} shows the inputs used to evaluate the end-to-end application.
All these inputs are from sequencing machines that have sequenced the human genome. Note the differences in the accuracy of the inputs, which refer to the errors introduced by the machines during the sequencing (reading) process.
ONT and PBCLR have an accuracy of 85\% and 88\%, respectively, while PBHF inputs have an accuracy of nearly 100\%.
PBHF inputs are obtained using ``PacBio High Fidelity" technology, which consists of reading the same piece of genome several times and mitigating the error by a consensus process.
This difference in accuracy is translated into different behavior during the read-mapping process.
A higher accuracy implies a lighter volume of work in the align stage when using SW.

We select the 18 most time-consuming sequences from each input set to keep simulation time in gem5 manageable. This allows us to reduce the execution time while maintaining the application's behavior.

\begin{table}[t]
\centering
\caption{Input sequence datasets.}
\begin{tabular}{@{}rccc@{}}
\toprule
 & \textbf{Sequencing machine} & \textbf{Avg. seq. length} & \textbf{Accuracy}  \\ \midrule
\textbf{ONT~\cite{hifiont}} & Oxford Nanopore & 17,710 & 85\% \\
\textbf{PB CLR~\cite{clr}} & PB Sequel II System & 6,739 & 88\% \\
\textbf{PB HF 1~\cite{hifiont}} & PacBio HiFi & 12,858 & 99.99\% \\
\textbf{PB HF 2~\cite{hifiont}} & PacBio HiFi & 15,602 & 99.99\% \\
\textbf{PB HF 3~\cite{hifiont}} & PacBio HiFi & 14,149 & 99.99\% \\ \bottomrule
\end{tabular}%
\label{tab:perf_inputs}
\end{table}



\section{Evaluation}\label{sec:evaluation}

In this section, first, we show how Squire can speed up the five evaluated kernels.
Then, we evaluate the impact the synchronization module has on the design by modifying the implementation to use software mutexes instead of Squire's hardware module.
We also evaluate the end-to-end read-mapper to understand how Squire improves a full application. Finally, we perform a design space exploration to justify the size of the caches used by the workers and perform an area and energy consumption study.

\subsection{Performance Evaluation}
\label{sec:performance_eval}

\begin{figure}[t]
    \centering
    \includegraphics[width=\columnwidth]{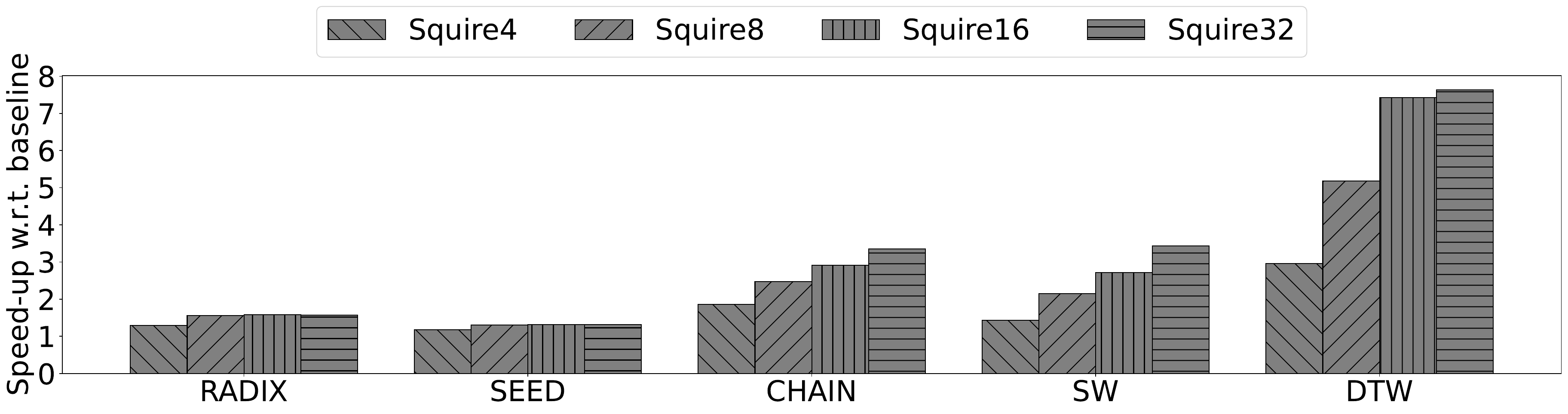}
    \caption{Squire evaluation for the five kernels described in Section~\ref{sec:methodology_kernels}. We evaluate Squire with 4, 8, 16, and 32 workers.}
    \label{fig:eval_kernels}
\end{figure}

Figure~\ref{fig:eval_kernels} shows the performance evaluation of Squire for the five kernels described in Section~\ref{sec:methodology_kernels} when changing the number of workers.

For RADIX and SEED, Squire achieves diminishing returns when using from 8 up to 32 workers, due to small input data size. As explained in Section~\ref{sec:radix_opt}, we stablish a minimum of 10,000 elements to use Squire. Below that, the initialization of Squire becomes the bottleneck in the sorting process. Maximum performance is achieved with 16 workers, reaching 1.58$\times$ for RADIX and 1.32$\times$ for SEED. 

Employing 32 workers for CHAIN and SW leads to noticeable speed-ups, unlike RADIX and SEED, reaching 3.35$\times$ and 3.43$\times$ with respect to the base system, respectively. The speedups from 16 to 32 workers are 1.19$\times$ and 1.26$\times$ for CHAIN and SW.

Finally, Squire obtains remarkable speedups up to 32 workers for DTW, reaching 7.64$\times$. However, we consider 16 workers the optimum point with a speedup of 7.42$\times$.

These results show that Squire can enable fine-grain parallelism on dependency-bond kernels. While Squire scales well with worker count if there is enough work to compute, we advocate that a balanced design should have between 8 and 16 workers. Doubling the number of workers to 32 does not compensate for the cost in the common case.

\subsection{Synchronization Module Evaluation}\label{sec:syncmod_eval}

\begin{figure}[t]
    \centering
    \includegraphics[width=0.7\columnwidth]{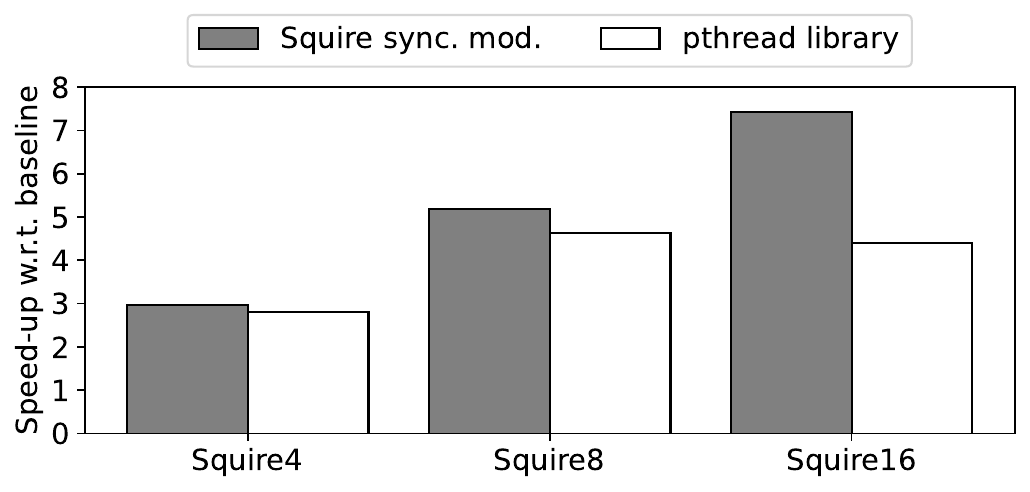}
    \caption{Squire performance evaluation when using the synchronization module vs the pthread library.}
    \label{fig:eval_swsem}
\end{figure}

Figure~\ref{fig:eval_swsem} shows the benefits of using the synchronization module in Squire for DTW kernel.
We show the results up to 16 workers since we have considered it the optimum point in Section~\ref{sec:performance_eval}.
We instantiate Squire without the synchronization module and synchronize through the \textit{pthread mutex} library.
We use DTW for this experiment since it is one of the kernels (along with SW) that uses the \textit{local counters}.

The synchronization module improves performance for any number of workers, increasing in importance as the number of workers increases.
We observe a speed-up of up 1.69$\times$ when using the synchronization module with 16 workers.

\subsection{End-to-End Application Evaluation} \label{sec:performance-evaluation}

\begin{figure}[t]
    \centering
    \includegraphics[width=\columnwidth]{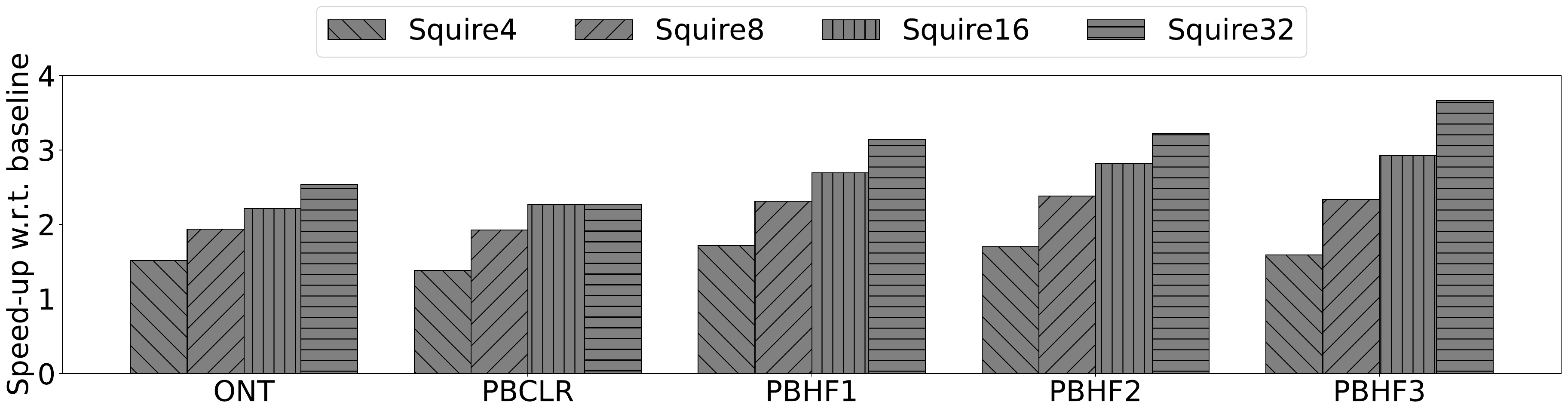}
    \caption{Squire evaluation for the end-to-end read-mapping application described in Section~\ref{sec:end2end_eval}.}
    \label{fig:eval}
\end{figure}

Figure~\ref{fig:eval} shows the performance evaluation of Squire for an end-to-end application for the five inputs described in Table~\ref{tab:perf_inputs}.
We evaluate Squire with 4, 8, 16, and 32 workers.
As stated in Section~\ref{sec:inputs}, the different datasets behave differently during the read mapping process; thus, the align stage has less weight for the PBHF inputs.

When looking at the whole read-mapping end-to-end application, Squire achieves speed-ups of up to 3.66$\times$.
For all the inputs, Squire scales well with worker count and accomplishes its best performance with 32 workers.
For ONT and PBCLR inputs, Squire achieves speed-ups of 2.54$\times$ and 2.27$\times$, respectively. For PBHF inputs, Squire achieves speed-ups higher than 3$\times$. A higher accuracy of the sequencing machines implies more work to process but smaller chunks of work, which favors Squire. As sequencing technologies keep improving, this trend will consolidate and devices like Squire will be more effective.

\subsection{Cache Size Exploration} \label{sec:cache-exploration}

Each worker has its own private L1 data and instruction caches, which will largely determine the area that Squire will occupy. For this reason, we perform a design space exploration study to make a judicious choice and minimize the area and power overhead of the design.

To evaluate the cache sizes, we use the end-to-end application and fix the number of workers to 16.
We use the ONT input dataset.
To evaluate the instruction cache size, we fixed the data cache size to 8 KB and vice versa. To measure performance, we use misses per kilo instructions (MPKI).

\begin{figure}[t]
    \centering
    \includegraphics[width=0.8\columnwidth]{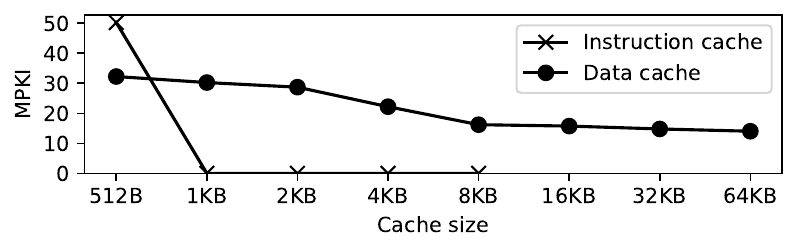}
    \caption{Misses per kilo instructions (MPKI) when changing the cache size for the instruction and data caches.}
    \label{fig:cache_eval}
\end{figure}

Figure~\ref{fig:cache_eval} shows MPKI when varying cache sizes.
For the instruction cache, we observe a drastic change when going from 512~B to 1~KB. Beyond that, MPKI remains close to zero. For the data cache, we see consistent improvement up to 8~KB, which we consider the sweet spot. A larger 16~KB data cache improves MPKI marginally at a large cost.
Therefore, we have employed 1~KB and 8~KB as instruction and data cache sizes, respectively, for all the experiments in this section.

\subsection{Area Overhead}

We use the Arm Neoverse N1 to model the out-of-order core. Using the public data for an N1~\cite{n1} at 7nm, the floor planned area is given as 1.15~mm$^{2}$. 

The workers we model could be compared to the Arm Cortex M35P microprocessor.
Using public data for an M35P at 40LP~\cite{m35p}, the floor planned area is given as 0.091~mm$^{2}$. This area already includes a 16~KB instruction cache.
The instruction cache included in the M35P is larger than the caches we employ since we employ 1~KB for L1I and 8~KB for L1D (see Section~\ref{sec:cache-exploration}).
Also, the M35P is a processor capable of booting an operating system, and our workers do not require as many functionalities as the M35P.
Therefore, we must consider that we are overestimating the area of the workers.

When employing 16 workers, the total area overhead at 40nm would be 1.456~mm$^{2}$.
To estimate the area with 7~nm, we scale these numbers, considering fin pitch, gate pitch, and interconnect pitch, using data from several studies~\cite{techcmp1,techcmp2,techcmp3,techcmp4,techcmp5,techcmp6} to arrive at a 12$\times$ area reduction when moving from 40~nm to 7~nm.
Thus, obtaining an area for a Squire component of 0.121~mm$^{2}$.

Therefore, we could place a 16-worker Squire component per core with an area overhead of 10.5\%.

\subsection{Energy Consumption}

\begin{figure}[t]
    \centering
    \includegraphics[width=\columnwidth]{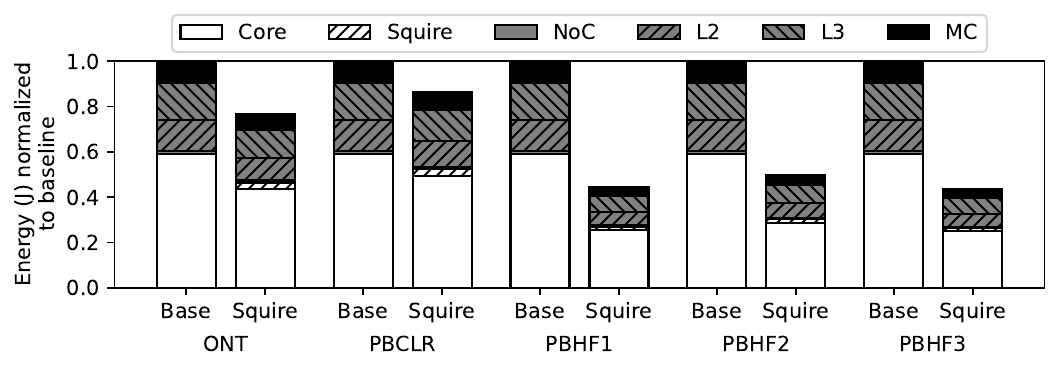}
    \caption{Energy consumption comparison between the baseline and when using Squire for the end-to-end application. 
    }
    \label{fig:power_eval}
\end{figure}

To estimate the Squire energy consumption, we use McPAT~1.3~\cite{mcpat} with the enhancements proposed by Xi et al.~\cite{mcpat2}.
We performed this estimation using a process technology node of 22 nm, a supply voltage of 0.8 V, and the default clock gating scheme.

Figure~\ref{fig:power_eval} shows the energy consumption of the baseline when using Squire with 16 workers for the end-to-end application.
The executions using Squire achieve significant energy reductions of up to 56\% over the baseline system for the PBHF3 input.
Similarly, Squire reduces consumption by 55\% and 50\% for PBHF1 and PBHF2 inputs.
The ONT and PBCLR inputs show a more modest energy reduction of 24\% and 14\%, respectively.


The host cores are the most energy-consuming components, followed by the L2 and L3 caches. The memory controllers and the NoC have a marginal energy consumption. The energy overhead introduced by Squire is small; we observe an energy overhead of around 6\% with respect to the host cores,
which is largely offset by the reduction in the rest of the components.

\section{Related Work}\label{sec:related}

We identify more general-purpose hardware accelerators like the Walkers~\cite{meetwalkers}, a programmable hardware accelerator for traversing hash tables in a database.
Transmuter~\cite{transmuter} and Versa~\cite{versa} propose a matrix of general-purpose processing elements interconnected by a mesh.
The accelerator is shared by all the cores of the chip.
The system can be reconfigured as a systolic array of processing elements, as a typical memory hierarchy, or as a private scratchpad for each processing element.
UPMEM~\cite{upmem} is the first publicly available general-purpose programmable PIM system.
AIM~\cite{aim-upmem} is a sequence alignment framework that uses UPMEM for the evaluation.

Table~\ref{tab:acc_cmp} shows a qualitative comparison between Squire and the other general-purpose hardware accelerators. 
The Walkers have a fixed pipeline, forcing the data to traverse it, thus limiting its flexibility.
In addition, the Walkers have a very limited ISA support and do not have any method for synchronization.
The compute units must execute the code completely in parallel without communicating with the rest.
Transmuter and Versa instantiate one accelerator shared for all the cores of a chip.
To exploit all the computing resources, the application should be split into two sets of threads, one that is executed on the accelerator and the other on the cores.
By contrast, Squire is a simpler private accelerator for each core.
The application is divided into as many threads as cores, and then each core performs nested parallelism in its Squire.
Moreover, the interconnection networks in Transmuter and Versa (among processing elements and between the accelerator and the host cores) add communication latencies with respect to Squire.
AIM is a processing in memory component that instantiates several processors per physical memory cell, thus losing the virtual memory capability and limiting the address range the processors can access.
Each processor controls a chunk of the memory and can not access the rest of the system.
To communicate with other processors, they must do it through main memory, which hinders performance~\cite{upmem-bible}.


A big.LITTLE architecture is composed of several cores with different design targets: computational performance and power efficiency~\cite{ARM_bigLITTLE}. All are capable of running system code and are visible to the operating system.
In contrast, Squire is a set of very simple cores with no system support and is subordinated to a host core.
Moreover, Squire is equipped with a synchronization module that allows for fast communication among its workers (see Section~\ref{sec:syncmod_eval}).

\begin{table}[t]
\caption{Qualitative comparison among several proposals.
}
\resizebox{\columnwidth}{!}{%
\begin{tabular}{@{}rccccc@{}}
\toprule
 & \textbf{Walkers\cite{meetwalkers}} & \textbf{Transmuter\cite{transmuter}} & \textbf{Versa\cite{versa}} & \textbf{AIM\cite{aim-upmem}} & \textbf{Squire} \\
\midrule
\textbf{Programable}                  & \checkmark  & \checkmark & \checkmark & \checkmark & \checkmark \\
\textbf{Rich ISA support}             & $\times$    & \checkmark & \checkmark & \checkmark & \checkmark \\
\textbf{Flexible datapath}            & $\times$    & \checkmark & \checkmark & \checkmark & \checkmark \\
\textbf{Virtual memory support}       & \checkmark  & \checkmark & \checkmark & $\times$   & \checkmark \\
\textbf{Rapid synchronization}        & $\times$    & \checkmark & \checkmark & $\times$   & \checkmark \\
\textbf{Private accelerator per core} & \checkmark  & $\times$   & $\times$   & $\times$   & \checkmark \\
\bottomrule
\end{tabular}%
}
\label{tab:acc_cmp}
\end{table}

\section{Conclusions}\label{sec:conclusions}

In this article, we propose Squire, a general-purpose accelerator for dependency-bound fine-grain parallelism.
Squire consists of a set of simple general-purpose in-order cores, called workers, and a synchronization module for rapid synchronization. Each host core is augmented with a Squire engine to offload fine-grain tasks.

We evaluate Squire on a simulated multicore SoC, obtaining speed-ups of up to 7.64$\times$ in dynamic programming kernels, and an acceleration for an end-to-end application of 3.66$\times$. 
We also evaluate the usage of resources and show that Squire achieves an energy reduction of up to 56\% with an area overhead of 10.5\% per core.

\section*{Acknowledgment}
This research was supported by MICIU/AEI/10.13039/501100011033 and by "ERDF A way of making Europe" through contracts [PID2023-146511NB-I00], [PID2023-146193OB-I00], and [PID2022-136454NB-C22]; by the Ministry for Digital Transformation and Public Service, (i) via the framework of the Recovery, Transformation and Resilience Plan - NextGenerationEU [REGAGE22e00058408992] and (ii) through the Càtedra Chip UPC project [grant number TSI-069100-2023-0015]; by the Generalitat of Catalunya through contract [2021-SGR-00763]; by the Arm-BSC Center of Excellence; by the Lenovo-BSC Framework Contract (2020); and by Government of Aragón (T58\_23R research group).  A. Armejach is a Serra Hunter Fellow.

\bibliographystyle{IEEEtran}
\bibliography{refs}

\end{document}